\documentstyle[prl,aps,preprint,epsfig,tighten]{revtex}

\begin{document}
\draft

\title{
A simple formula for the L-gap width of a face-centered-cubic
photonic crystal
}

\author{
Alexander Moroz\thanks{http://www.amolf.nl/external/wwwlab/atoms/theory/} 
}

\address{FOM-Instituut voor  Atoom- en Molecuulfysica, Kruislaan 407, 
1098 SJ Amsterdam,  The Netherlands
}
\address{
\begin{abstract}
The width $\triangle_L$ of the first Bragg's scattering peak in the (111)
direction of a face-centered-cubic lattice of air spheres can be well 
approximated by a simple formula which only involves the volume averaged
$\varepsilon$ and  $\varepsilon^2$ over the
lattice unit cell, $\varepsilon$
being the (position dependent) dielectric constant of the medium,  and 
the effective
dielectric constant $\varepsilon_{e\!f\!f}$ in the 
long-wavelength limit approximated by Maxwell-Garnett's formula. 
Apparently, our formula describes the asymptotic behaviour 
of the absolute gap width  $\triangle_L$ for high dielectric contrast
$\delta$ exactly. The standard deviation $\sigma$
steadily decreases well below $1\%$ as $\delta$ increases. 
For example $\sigma< 0.1\%$ for the sphere filling fraction $f=0.2$ 
and $\delta\geq 20$. 
On the interval $\delta\in(1,100)$, our formula 
still approximates gap widths 
with a reasonable precision, namely 
the absolute gap width $\triangle_L$ 
with a standard deviation  $3\%$  for low filling fractions
up to $6.5\%$ for the close-packed case and the relative gap width 
$\triangle_L^r$ from $4.2\%$ to $8\%$.
Differences between  the case of air spheres in a dielectric and 
dielectric spheres in air are briefly discussed.\vspace*{1.4cm}\\
\end{abstract}
}


\address{
({\bf J. Opt. A: Pure Appl. Opt. 1, 471-475 (1999)})
\vspace*{1.4cm}}
\maketitle


\newpage
\narrowtext
\section{Introduction}
The propagation of light in a periodic dielectric medium has recently
attracted much attention due to the possibility of opening 
a gap in the spectrum of  electromagnetic waves
for both polarizations and all directions of the incident waves 
\cite{Y,Jo,LL,HCS,YGL}. In such a 
medium, the density of states (DOS) can be, in a certain frequency interval, 
either reduced down to zero (photonic band gap) or enhanced with respect 
to the vacuum case. The changes in the DOS affect various physical quantities.
The most transparent is the change in the spontaneous emission rate
of embedded atoms and molecules which may have applications for 
semiconductor lasers, heterojunction bipolar transistors, 
and thresholdless lasers \cite{Y} or  to  create  new sources of 
light for ultra-fast optical communication systems.

Existence of the full photonic band gap was first demonstrated at 
microwaves \cite{YGL}. Recently, thanks to the intense experimental effort, 
we have witnessed a significant progress in fabrication of 
complete photonic-bandgap structures at near-infrared \cite{KRB,Lin,Lev}.
In two dimensions,  complete photonic-bandgap structures 
have been fabricating for TM polarization \cite{KRB}.
In three dimensions, a promissing structure has been fabricated
by Sandia's group \cite{Lin}. Nevertheless, in one direction this structure 
extends less than two unit cells and 
there is ongoing experimental search to improve its properties.
One of the most promising candidates to achieve a complete 
photonic bandgap at  optical wavelengths and fabricate large enough
structures at near-infrared wavelengths are collodial systems of 
microspheres. Indeed, the latter can self-assemble into three-dimensional 
crystals with excellent long-range periodicity with the lattice
constant well  below infrared scale \cite{Pier,Bog,TW,PBW,BRW}.
This long-range periodicity gives rise to strong optical
Bragg's scattering clearly visible by the naked eye and already described 
in 1963 \cite{LKW}.
Monodisperse collodial suspensions of microspheres crystalize either 
in a face-centered-cubic (fcc) \cite{TW,As,WV} or (for small
sphere filling fraction) in a
body-centered-cubic (bcc) lattice \cite{PBW}. 
Using suspensions of microspheres of different sizes one can also 
prepare crystals with a complex unit cell (containing more 
than one scatterer). Both the case of ``dense spheres'' \cite{WV} and 
``air spheres'' \cite{IP}
when the dielectric constant of spheres $\varepsilon_s$ is greater 
and smaller  than the dielectric 
constant $\varepsilon_b$ of the background medium,
respectively,  can be realized experimentally.
There is a significant difference between the two cases, since,
according to numerical calculations, simple dielectric lattices of 
homogeneous spheres \cite{SHI,BSS,MS} in air do 
not exhibit a full photonic band gap, while for air spheres
a full band gap {\em can} open for a simple fcc lattice 
\cite{SHI,BSS,MS}. 
Unfortunately, the required dielectric contrast 
$\delta=\max (\varepsilon_s/\varepsilon_b,\varepsilon_b/\varepsilon_s)$ 
for opening the full band gap, 
 either $\stackrel{\textstyle >}{\sim}8.4$ obtained using the plane 
wave method  \cite{BSS}, or, $\stackrel{\textstyle >}{\sim}8.13$  obtained 
by the photonic analoque of the  Korringa-Kohn-Rostocker (KKR)  
method \cite{MS,Mo},  is currently out of experimental reach at optical and 
near-infrared frequencies for photonic colloidal structures. 
The absence of a full gap in this 
frequency range in currently available collodial crystals of homogeneous
and single size spheres
does not mean the absence of interesting physics in this
weak photonic region.
For example,  the change in the spontaneous 
emission rate of dye molecules in an fcc collodial crystal can be 
observed already at a relatively low 
$\delta\approx 1.2$ \cite{PBK}.

In contrast to the full gap, Bragg's reflection can be observed 
for arbitrarily  small $\delta$ as long as a sample has 
sufficient long-range periodicity. Analysis of  
Bragg's scattering might not only be  useful to understand 
the physics of photonic crystals, but it has already found 
practical application in distributed feedback 
lasers in the visible region of the spectrum \cite{ML}.
The first Bragg's peak can be characterized by the width of 
the (lowest) stop gap
(gap at a fixed direction of the incident light) at a
certain point on the boundary of the Brillouin zone. 
We focus here on the case of a simple
fcc lattice of air spheres \cite{IP}, which is among the most
promising candidates to achieve a full photonic band gap.
For an fcc lattice, it is convenient to consider Bragg's scattering in  
the (111) direction which corresponds to the L direction of the 
Brillouin zone
(see \cite{Kos} for the classification of special
points of three-dimensional lattices). Apart from numerous experimental 
data now available \cite{TW,As,WV,YG}, there are at least two other
reasons for this choice. First, the width of the first stop gap takes on 
its  maximum at the L point and, second, experimental techniques 
make it possible to allow one
to grow collodial crystals such that the L direction corresponds
to normal incidence on the crystal surface.

Let $\varepsilon({\bf r})$ be the dielectric constant of an fcc 
photonic crystal. One has $\varepsilon({\bf r})=\varepsilon_s$ 
if ${\bf r}$ is inside the sphere and $\varepsilon({\bf r})=\varepsilon_b$ otherwise.
Let $f$ be the sphere filling fraction, i.e.,
volume of the sphere(s) in the unit cell per unit cell volume.
Once $f$ is fixed, the spectrum is only a function of the dielectric 
contrast $\delta$. By a suitable rescaling, one can always set 
$\varepsilon_s=1$ 
for the case of ``air'' spheres ($\varepsilon_b=1$ for the case of 
``dense'' spheres). 
As $\delta$ and $f$ are varied, both the absolute L-gap 
width $\triangle_L$ and the L-midgap frequency $\nu_c$ change. 
As a function of $\delta$, $\triangle_L(\delta)$
takes on its maximum at some $\delta=\delta_m(f)$ while
$\nu_c(\delta)$ monotonically decreases.
We address the question of whether the width $\triangle_L$
can be understood in terms of simple quantities, 
namely, the volume averaged dielectric constant, 
$$\overline{\varepsilon} =
f\varepsilon_s +(1-f)\varepsilon_b,
$$
the volume averaged $\varepsilon^2({\bf r})$,
$$\overline{\varepsilon^2}=
[f\varepsilon_s^2 +(1-f)\varepsilon_b^2],
$$
and the effective dielectric constant $\varepsilon_{e\!f\!f}$.
The latter characterizes optical properties of the crystal 
in the long-wavelength limit and is (theoretically)
determined by the slope of the
linear part of the band structure,
$\varepsilon_{e\!f\!f}^{-1/2} =\lim_{k\rightarrow 0}\,  d\omega/(c dk)$.
Note that due to the vector character of electromagnetic waves, 
 $\varepsilon_{e\!f\!f}$ differs from $\overline{\varepsilon}$,
in contrast to the scalar case where
$\varepsilon_{e\!f\!f}=\overline{\varepsilon}$ \cite{DCH}. 
One can show that for any $f$, $\varepsilon_b$, and $\varepsilon_s$,
\begin{equation}
\varepsilon_{e\!f\!f}\leq  \overline{\varepsilon} \leq 
\sqrt{\overline{\varepsilon^2}}.
\label{ineq}
\end{equation}
Equality in (\ref{ineq}) occurs if and only if either $f=0$ or $f=1$,
or, if $\delta=1$, i.e., if $\varepsilon_b=\varepsilon_s$.
The effective dielectric constant can be well approximated 
\cite{MS,DCH} by Maxwell-Garnett's formula \cite{MG},
\begin{equation}
\varepsilon_{e\!f\!f} \approx 
\varepsilon_{e\!f\!f}^{MG} =\varepsilon_b \,
(1+ 2\, f\alpha)/(1- f\alpha),
\label{maxgsp}
\end{equation}
where, for a homogeneous sphere, 
the polarizability factor
$\alpha=(\varepsilon_s-\varepsilon_b)/(\varepsilon_s+2\varepsilon_b)$.
Note, however, that in the case of air (dense) spheres
$\varepsilon_{e\!f\!f}^{MG}$ slightly 
{\em overestimates} ({\em underestimates})
the exact value of $\varepsilon_{e\!f\!f}$ as calculated from the
band structure \cite{MS,DCH}.

\section{Results}
\begin{figure}[tbp]
\begin{center}
\epsfig{file=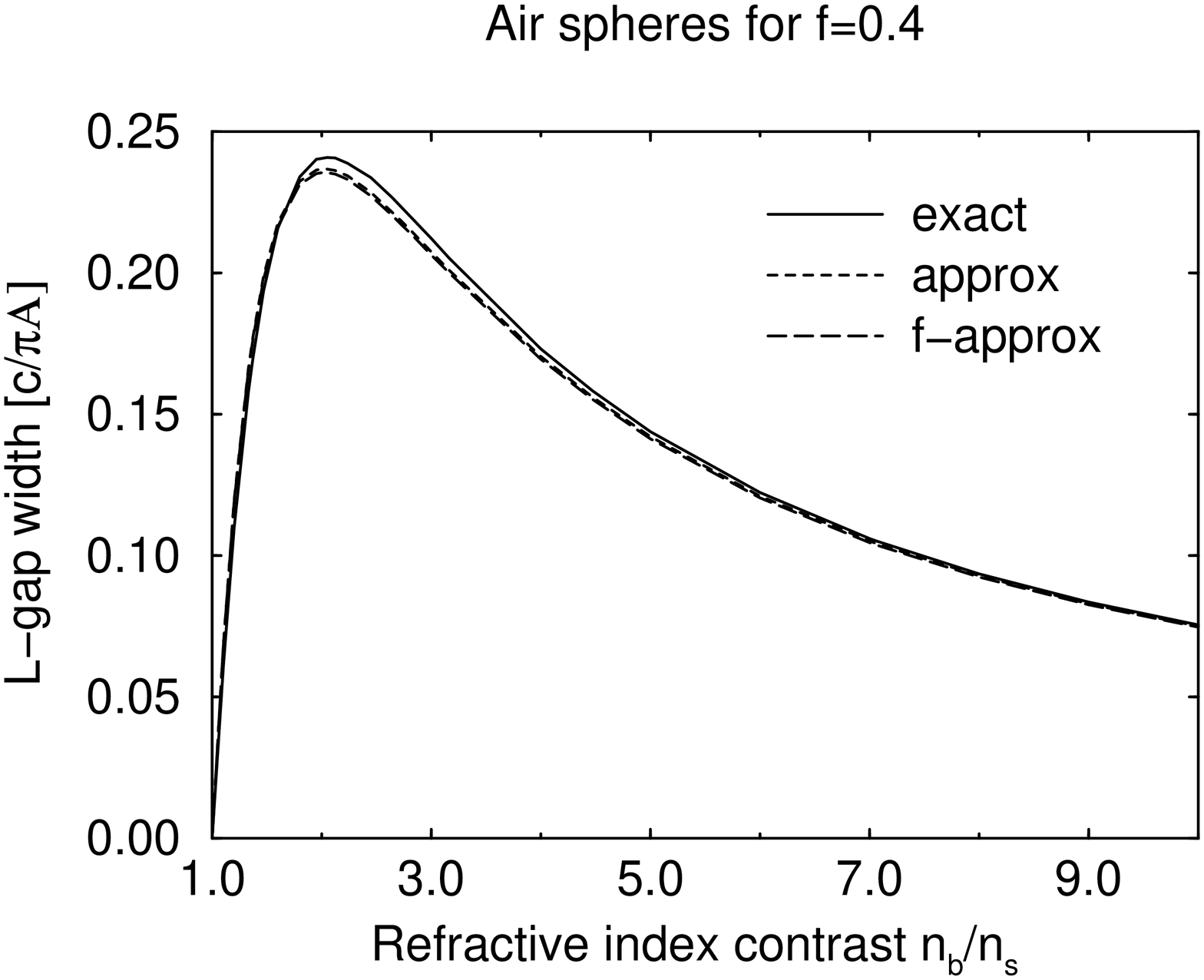,width=10cm,clip=0,angle=-90}
\end{center}
\caption{The L-gap width in units
$c/\pi A$, where $c$ is the speed of light in vacuum and $A$
is a length of the side of the conventional unit cell of the 
cubic lattice [25],
for an fcc lattice of air spheres 
with $f=0.4$ calculated exactly (the solid line)
and approximated by the formula (\ref{best})
with optimized $C$ (the dashed line) and  with 
$C$ given by the formula (\ref{cf})
(the long-dashed line). The last two curves almost
overlap. 
}
\label{aproxff}
\end{figure}

Obtaining exact analytic results for dielectric lattices turns out to 
be notoriously difficult and 
numerics has been the main tool to understand photonic gaps
so far \cite{LL,HCS,SHI,BSS,MS,DCH}.
A simple analytical formula, if any, may be a good starting point for 
obtaining a better insight into the problem.
It was rather surprising to find out that such a formula 
can be found for the L-gap width $\triangle_L$.
Namely, in the case for air spheres, $\triangle_L$ can 
be approximated by the formula
(see Fig.\ \ref{aproxff})
\begin{equation}
\triangle_L \approx C g = \left. C \left(\sqrt{
    \overline{\varepsilon^2}} -\varepsilon_{e\!f\!f}\right)^{1/2}
\right/ \bar{\varepsilon}.
\label{best}
\end{equation}
For a given filling fraction $f$, the constant $C=C(f)$  
was determined by taking the average over 
$\triangle_L/g$ where $\triangle_L$ is 
the L-gap width calculated numerically using  
a photonic analogue \cite{Mo} of the KKR method \cite{KKR}.
The latter method gives results which are
in excellent agreement with experimental values \cite{WV}.

Apparently, for sufficiently high $\delta\gg\delta_m(f)$, our 
formula captures the asymptotic behaviour 
of the absolute gap width  $\triangle_L$ exactly. 
The standard deviation $\sigma$
steadily decreases well below $1\%$ as 
one investigates region $\delta\geq \delta_c$ for higher
and higher $\delta_c$. 
For $\delta\geq 20$ one has  $\sigma< 0.1\%$ for $f=0.2$,
$\sigma< 0.4\%$ for $f=0.1$, and
$\sigma< 0.3\%$ for $f=0.4$ (see, for example, Fig.\ \ref{aproxff}).
For $\delta\geq 36$ and $f=0.6$, $\sigma< 0.5\%$,
while for the close-packed case, $\sigma\leq 1\%$.
If, however, in the latter case  $\delta\geq 50$, 
$\sigma$ drops below $0.7\%$.

For $\delta\in(1,100]$, our formula (\ref{best})
still describes  $\triangle_L$ with a reasonable
accuracy ranging from $3.3\%$ to $6.5\%$
(depending on the filling fraction).
The values of $C$, their standard quadratic deviation $\sigma$,
and the relative error $\sigma_r=\sigma/C$
are shown in Tab. I.
Approximately thirty values of the dielectric contrast 
within the interval $\delta\in(1,100]$ were
taken for every filling fraction considered.
For a given filling fraction, the main part of the error is picked up
\begin{center}
TABLE I. The values of $C$, 
their standard quadratic deviation $\sigma$,
and the relative error $\sigma_r=\sigma/C$
for different filling fractions and $\delta\in(1,100]$. 
\vspace*{0.2cm} \\
\begin{tabular} {|c|c|c|c|c|c|} \hline\hline
 &  $f=0.1$ & $f=0.2$ & $f=0.4$ & $f=0.6$ & $f=0.74$      \\
\hline
 $C$  & 0.762  & 0.868 & 0.875 & 0.808 &  0.736   \\
\hline 
 $\sigma$  &  $0.031$   & $0.03$ & $0.029$ &  $0.038$ &  $0.048$    \\
\hline 
 $\sigma_r$  &  $4.1\%$   & $3.4\%$ & $3.3\%$ &  $4.7\%$ &  $6.5\%$ \\
 \hline\hline
\end{tabular}
\end{center}
\vspace*{0.2cm} 
%
\noindent
around $\delta=\delta_m(f)$ for which $\triangle_L$
takes on its maximum. At the maximum of 
the L-gap width $\triangle_L$ our formula (\ref{best})
gives persistently a slightly lower value for $\triangle_L$. 
Note that for moderate $\delta$ 
Maxwell-Garnett overestimates $\varepsilon_{e\!f\!f}$
for the case of air spheres \cite{MS,DCH}. Therefore, using the exact
$\varepsilon_{e\!f\!f}$ may reduce errors further.

According to Tab. I, the quantity $C$ shows a weak dependence on $f$
which can be  approximated with high accuracy (relative error 2.5\%) 
by the formula
\begin{equation}
C(f)=C_{0} + 0.14\, f\,(2 f_{m} -f)/f_{m}^2.
\label{cf}
\end{equation}
Here $C_0\approx 0.74$ is  the minimal value of $C$
and $f_m$ is the filling fraction for which 
$C$ takes on its maximal value. Table I
indicates that $C$ takes on its minimal value $C_0$ at the extreme filling
fractions $f=0$ and $f=0.74$, and its maximal value is $C_m\approx 0.88$ at 
$f_m\approx 0.74/2$. The factor $0.14$ in the interpolation 
formula (\ref{cf}) is the difference $C_m-C_0$.
Using $C(f)$ in (\ref{best}) does not raise the relative error 
$\sigma_r$
more than $0.4\%$ for intermediate filling fractions.
Fig.\ \ref{aproxff} shows approximations to the 
 L-gap width for $f=0.4$ using formula (\ref{best}) with optimized
$C$ taken from Tab. I and with $C(f)$  given by the formula (\ref{cf}).
As the dielectric contrast $\delta$ increases,
$\triangle_L$ first increases to its maximal value and then
slowly decreases as $\delta^{-1/2}$. This behavior
is well reflected by our formula (\ref{best}) which in the limit
$\delta\gg 1$ yields
\begin{equation}
\triangle_L \sim  C(f) 
\frac{2}{\sqrt{(1-f)(2+f)}}
\left(\frac{2+f}{2\sqrt{1-f}} - 1\right)^{1/2}\, \delta^{-1/2}.
\end{equation}

\begin{figure}[tbp]
\begin{center}
\epsfig{file=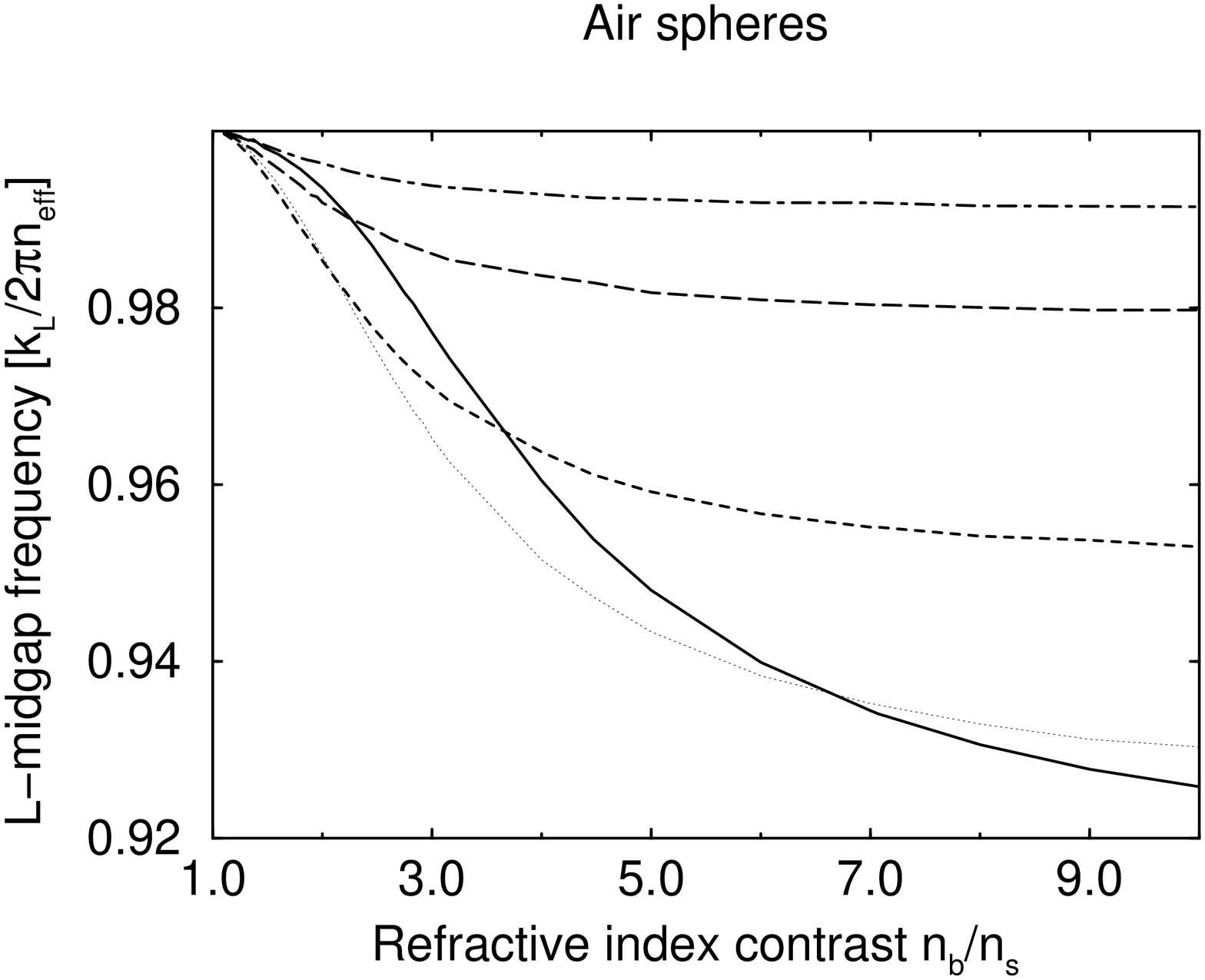,width=10cm,clip=0,angle=-90}
\end{center}
\caption{
The L-midgap frequency $\nu_c$ for air spheres in the units
$k_L/(2\pi n_{e\!f\!f}^{MG})$, where $k_L$ is the length of the Bloch vector
at the L point and $n_{e\!f\!f}^{MG}$ is the effective refractive index
of the medium calculated by Maxwell-Garnett's formula (\ref{maxgsp}).
The dot-dashed line is for the sphere filling fraction $f=0.1$, the 
long-dashed line is for $f=0.2$, the dashed line is for $f=0.4$, 
finely dotted line is for $f=0.6$, and the solid line corresponds to 
the close-packed case ($f=0.74$). 
}
\label{lgc}
\end{figure}

\begin{figure}[tbp]
\begin{center}
\epsfig{file=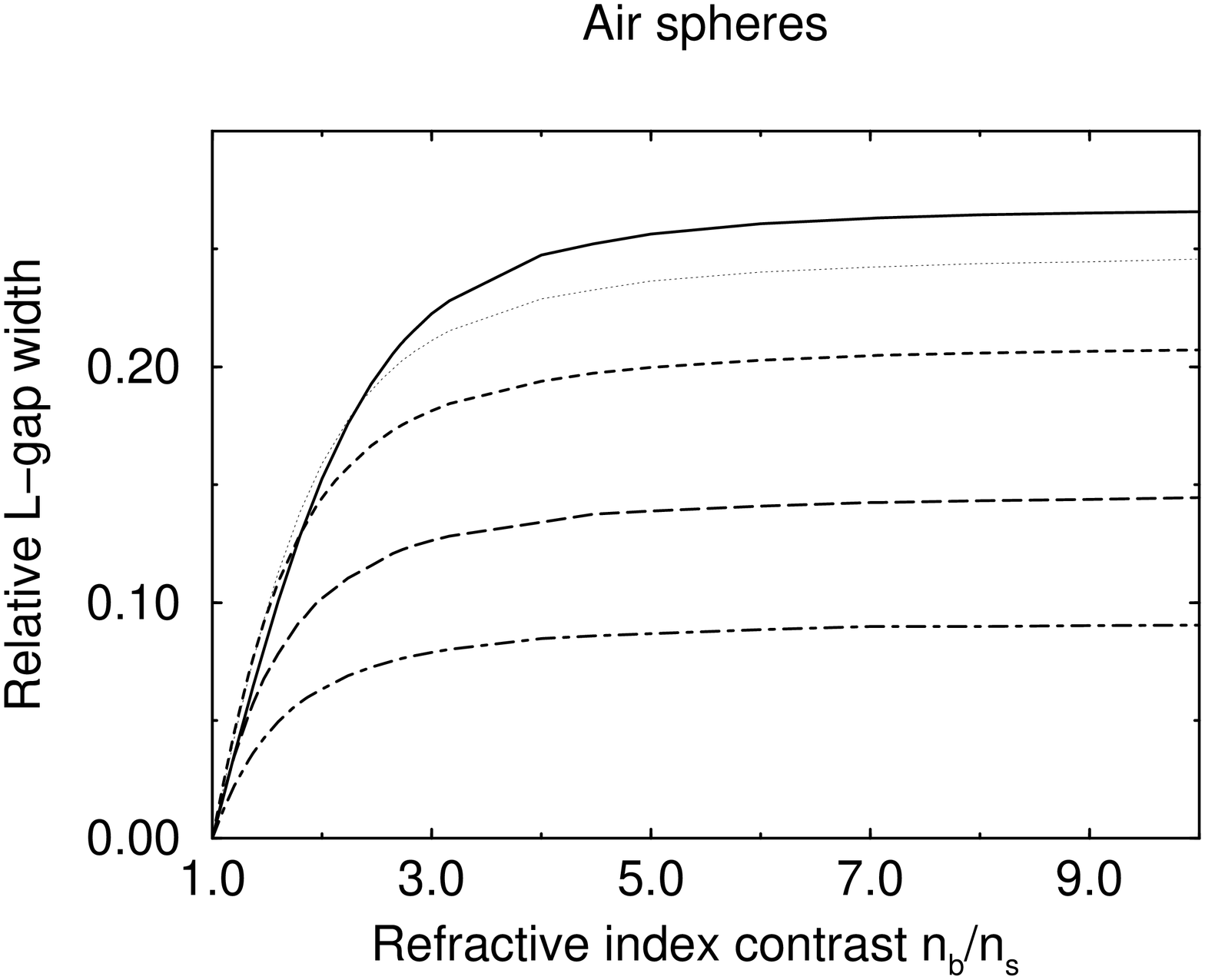,width=10cm,clip=0,angle=-90}
\end{center}
\caption{The relative L-gap width (the L-gap width divided by the
midgap frequency) for an fcc lattice of air spheres shows a rather simple 
dependence on the refractive index contrast 
$n_b/n_s=\sqrt\protect{\delta\protect}$.
The dot-dashed line is for the sphere filling fraction $f=0.1$, the 
long-dashed line is for $f=0.2$, the dashed line is for $f=0.4$, 
finely  dotted line 
is for $f=0.6$, and the solid line corresponds to the close-packed case 
($f=0.74$). 
}
\label{rlgw}
\end{figure}

\begin{figure}[tbp]
\begin{center}
\epsfig{file=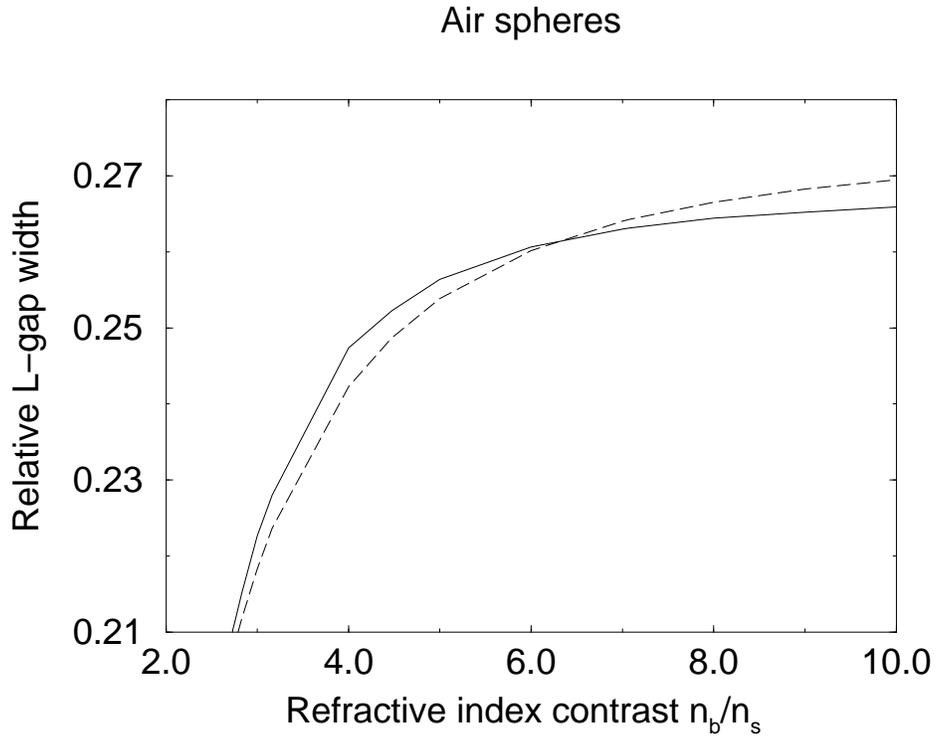,width=10cm,clip=0,angle=-90}
\end{center}
\caption{The relative L-gap width $\triangle^r_L$ for 
the close-packed fcc lattice 
of air spheres (solid line) and its approximation 
using the formula (\ref{relg})  with optimized $\overline{R}$ 
(long-dashed line). 
Once $\overline{R}$ is optimized, $\tilde{\sigma}_r$ can be reduced 
to  $\approx 1\%$. Even at $\delta=100$, the error in 
$\triangle^r_L$ is less than $2\%$. 
}
\label{rlcpap}
\end{figure}

Since the L-midgap frequency $\nu_c$ changes as 
$f$ and $\delta$ are varied, an invariant characteristic of Bragg's 
scattering is provided by the
relative L-gap width 
$$\triangle^r_L=\triangle_L/\nu_c.$$ 
$\triangle^r_L$ increases  monotonically 
as  $\delta$ increases (see Fig.\ \ref{rlgw})
and saturates very fast for $\delta>\delta_m$. Our observation here 
is that the L-midgap frequency $\nu_c$ can be well approximated by
\begin{equation}
\nu_c \approx c k_L/(2\pi n_{e\!f\!f}^{MG}),
\label{nucap}
\end{equation}
where $n_{e\!f\!f}^{MG}=\sqrt{\varepsilon_{e\!f\!f}^{MG}}$ and
$k_L$ is the length of the Bloch vector
at the L point. In units where the length of the side of 
the conventional unit cell of the cubic lattice \cite{AsM} 
is $A=2$, one has $k_L/\pi = \sqrt{0.75}$.
Recent measurements of $\nu_c$ for moderate $\delta$ \cite{WV} 
agree well with formula (\ref{nucap}) (see also Fig.\ \ref{lgc}).
For all filling fractions considered $\delta$ within the range
and $1\leq \delta\leq 100$, the  maximal deviation of the L-midgap 
frequency given by formula (\ref{nucap}) is less than
$8\%$ with respect to the exact value. 
Therefore, the formula 
\begin{equation}
\triangle^r_L \approx 2 \pi n_{e\!f\!f} ^{MG}\triangle_L/ k_L
\label{relg}
\end{equation}
is a natural candidate to describe  $\triangle^r_L$.
However, as shown in Fig.\ \ref{lgc},  formula (\ref{nucap}) 
systematically overestimates the exact value of $\nu_c$ by 
a little bit. This systematic error is also apparent 
from Fig.\ \ref{rlcpap}.
Due to the systematic error, the relative L-gap width $\triangle^r_L$ 
is described by the formula (\ref{relg}) with a slightly larger
relative error than is $\triangle_L$ by the formula (\ref{best}).
There are now two main contributions to the errors, one around 
the maximum of $\triangle_L$ and the other due to the systematic error.  
However, in the asymptotic region $\delta\gg \delta_m$
the first contribution disappears
whereas the systematic error saturates (see  Fig.\ \ref{lgc}).
As a result, in the asymptotic region, the relative error 
$\tilde{\sigma}_r$  is still within $\approx 1\%$. Fig.\ \ref{rlcpap}
shows that even at $\delta=100$ the error in $\triangle^r_L$
is less than $2\%$.

For $\delta\in(1,100]$, our formula (\ref{best})
still describes  $\triangle_L$ with a reasonable
accuracy 
ranging from 4.1\% to 8\% (depending on the filling
fraction) 
For a given filling fraction, the relative error 
$\tilde{\sigma}_r=\tilde{\sigma}/\overline{R}$
was determined by calculating the standard quadratic deviation 
$\tilde{\sigma}$ of the average value $\overline{R}$ 
of the ratio 
$R=\triangle^{r;exact}_L/\triangle^{r;approx}_L$,
where $\triangle^{r;exact}_L$ is the exact value of 
$\triangle^r_L$ calculated
numerically and  $\triangle^{r;approx}_L$ is its approximation 
calculated using
Eq. (\ref{relg}). The values of $\overline{R}$, $\tilde{\sigma}$,
and  $\tilde{\sigma}_r$ are collected in Table II. 
\begin{center}
TABLE II. The average value $\overline{R}$,
the standard quadratic deviation $\tilde{\sigma}$, and
the relative error $\tilde{\sigma}_r$ for $\triangle^r_L$
approximated by Eq. (\ref{relg})
for different filling fractions and $\delta\in(1,100]$. 
\vspace*{0.2cm} \\
\begin{tabular} {|c|c|c|c|c|c|} \hline\hline
 &  $f=0.1$ & $f=0.2$ & $f=0.4$ & $f=0.6$ & $f=0.74$     \\
 \hline 
 $\overline{R}$  &  $1.004$   & $1.012$ & $1.025$ &  $1.030$ &  
  $1.031$  \\
\hline
 $\tilde{\sigma}$  &  $0.044$   & $0.042$ & $0.052$ &  $0.068$ &  $0.082$  \\
\hline 
 $\tilde{\sigma}_r$  &  $4.37\%$   & $4.17\%$ & $5.05\%$ &  $6.61\%$ &  $7.95\%$ \\
 \hline\hline
\end{tabular}
\end{center}
\vspace*{0.2cm} 
%
\noindent
One expects a deviations
[of the order $5\%$ from the behavior described by the 
formula (\ref{relg})]
only in a rare case when a Mie resonance crosses the edge 
of the L-gap \cite{MT}.

\section{Discussion}
Formulas (\ref{best})  [together with (\ref{cf})], (\ref{nucap}), 
and (\ref{relg}) are the main results of this work.
They fit nicely experimental data on Bragg's
scattering in fcc photonic crystals of air spheres \cite{As,WV,IP}.
Note that $\triangle_L$ also characterizes the transmission of light
through such a crystal (see \cite{YG} for microwaves). 
The fact that such simple relations can describe one of
the photonic gaps has been completely unexpected.
Indeed, the numerical calculation of photonic band structures is 
a great deal more involved than that in the case of scalar waves
(including the case of electrons) where no analog of formulas 
(\ref{best}) and (\ref{relg}) is known.
Numerics has been the main tool to understand photonic gaps
\cite{LL,HCS,SHI,BSS,MS,DCH}. This is also the case of
two recent discussions of Bragg's scattering in the
(111) direction \cite{SY1}.
A previous attempt to understand Bragg's scattering in photonic crystals
involved an introduction of a ``photonic strength'' 
parameter $\Psi=3f\alpha$ \cite{WV}.
It was shown that the dynamical diffraction theory \cite{Zach},
which is well known in x-ray diffraction, already fails to describe
Bragg's scattering in a photonic crystal  for
$\Psi\approx 0.5$ \cite{WV}.

Formulas (\ref{best}) and (\ref{relg}) immediately raise
questions whether one can understand and derive them analytically.
The L-gap width for fcc structures is a natural measure 
to characterize their scattering strength, because, in contrast
to the full band gap, $\triangle_L\neq 0$ for arbitrarily small
$f$ and $\delta$. The latter suggests to take $\triangle^r_L$ given 
by Eq. (\ref{relg}) as a natural ``photonic strength'' parameter for
the air spheres case.
Neither the parameter $\Psi$ \cite{WV}, nor the parameter
$
\varepsilon_r = \left. 
\left(\overline{\varepsilon^2} - \overline{\varepsilon}^2\right)^{1/2}
\right/ \overline{\varepsilon},
$
introduced in \cite{SHI}, are directly related to a gap width.
However,
it turns out that formulas (\ref{best}) and (\ref{relg}) 
cannot be applied to the case of dense spheres. 
The simple fcc lattices of air and dense spheres have for the same 
dielectric contrast rather different behavior
with respect to the full photonic band gap \cite{SHI,BSS,MS}
and to the first Bragg's peak \cite{TW,As,WV,IP}. 
Our numerical calculation 
shows that, for dense spheres, $\triangle^r_L$ does not increase
monotonically with $\delta$ as in the case of air spheres.
Instead $\triangle^r_L$ first reaches a local maximum, then 
it returns to zero and only afterwards starts to increase
monotonically \cite{Mp}.
This behavior is reminiscent to that of the relative 
X-gap width $\triangle^r_X$
(X is another special point of the Brillouin zone of an fcc lattice
\cite{Kos}) in the case of air spheres \cite{LL,YG}.
It has been argued that the vanishing of $\triangle^r_X$
is due to the vanishing of the scattering form-factors \cite{LL,YG}.
Also, if $\triangle^r_L$ is plotted against the filling fraction,
one observes that the maximum of $\triangle^r_L$ shifts
to lower $f$ for dense spheres and towards close-packing for
air spheres \cite{Mp}.
It would be interesting to understand what causes this
different behavior. The latter can be partially attributed to the 
fact that, 
for a lattice of spheres, $\varepsilon_b$ no 
longer describes the dielectric constant of the surrounding medium, 
which
is instead described by the effective dielectric constant
$\varepsilon_{e\!f\!f}$. Therefore, the bare dielectric contrast 
$\delta$ 
is renormalized to 
$\delta_{e\!f\!f}=\max (\varepsilon_s/\varepsilon_{e\!f\!f},
\varepsilon_{e\!f\!f}/\varepsilon_s)$, where 
$1 < \varepsilon_{e\!f\!f} < \delta$ for  
$\varepsilon_s\neq \varepsilon_b$.
Given the bare dielectric contrast $\delta$, one finds that
the  renormalized dielectric contrast 
$\delta^d_{e\!f\!f}=\varepsilon_s/\varepsilon_{e\!f\!f}$ in the case 
of dense spheres is 
always smaller than the  renormalized dielectric contrast 
$\delta^a_{e\!f\!f}=\varepsilon_{e\!f\!f}$
in the case of air spheres \cite{MS}.
The latter is easy to verify in the limit when the
bare dielectric contrast $\delta$ tends to infinity, 
where the Maxwell-Garnett equation (\ref{maxgsp}) implies
\begin{equation}
\delta_{e\!f\!f}^d \sim \delta \,(1-f)/(1+2f)
< \delta_{e\!f\!f}^a \sim \delta\,(1-f)/(1+f/2).
\end{equation}
Nevertheless, a full understanding of the differences between the
lattices of air and dense spheres still remains a theoretical
challenge.

\section{Conclusion}
To conclude we have found that, despite of the complexity 
of the problem of propagation of electromagnetic waves in a 
periodic dielectric medium, the absolute and the relative 
width of the first Bragg's peak in the (111) direction for 
an fcc lattice of 
air spheres can be accurately described by the simple 
empirical formulas (\ref{best}) and (\ref{relg}), respectively. 
Apparently, for sufficiently high $\delta\gg 1$,  
our formula (\ref{best}) 
captures the asymptotic behaviour of $\triangle_L$ exactly.
Indeed, the relative error $\sigma_r$
steadily decreases as one investigates region 
$\delta\geq \delta_c$ for higher and higher $\delta_c$.   
For all filling fractions $\sigma_r$ falls 
well below $1\%$ if sufficciently high $\delta$
is taken. For example for  $\delta\geq 20$  one obtains 
$\sigma_r< 0.1\%$ for $f=0.2$ and  $\sigma_r< 0.3\%$ for $f=0.4$.
For $\delta\in(1,100]$  formula (\ref{best}) still describes  
$\triangle_L$ with a reasonable precision, namely, with the
relative error ranging from 3.3\% to 6.5\% (depending on the filling
fraction). The main contribution
to the error is picked up  
around $\delta=\delta_m(f)$ for which $\triangle_L$ takes on its maximum.
At $\delta=\delta_m$, our formula (\ref{best})
gives persistently a slightly lower value for $\triangle_L$. 
The relative L-gap width $\triangle^r_L$ is described
by the formula (\ref{relg}) with a slightly larger
relative error ranging from 4.1\% to 8\% (depending on the filling
fraction). The reason is that there are now two main contributions
to the error, that around the maximum of $\triangle_L$
and the second systematic error due to
the overestimation of the L-midgap frequency $\nu_c$ 
when using Eq. (\ref{nucap}).
All the formulas only involve the effective dielectric constant of 
the medium $\varepsilon_{e\!f\!f}$
approximated  by Maxwell-Garnett's formula (\ref{maxgsp}), and 
volume averaged $\varepsilon({\bf r})$ and $\varepsilon^2({\bf r})$
over the lattice unit cell. Since  $\overline{\varepsilon}$, 
$\overline{\varepsilon^2}$, and $\varepsilon_{e\!f\!f}$ have 
well-defined meaning for any lattice, this suggests that a similar gap 
behavior may occur for other lattices.
It would be interesting to find out if the same is true
for the width of the full photonic band gap \cite{Mp}.

I would like to thank A. van Blaaderen, A. Tip, and W. L. Vos
for careful reading of the manuscript and useful comments,
and other members of the photonic crystals interest group for discussion.
This work is part of the research program by  the Stichting voor Fundamenteel 
Onderzoek der Materie  (Foundation for Fundamental Research on
Matter) which  was made possible by financial support 
from the Nederlandse
Organisatie voor Wetenschappelijk Onderzoek (Netherlands Organization for
Scientific Research).  SARA computer facilities are also gratefully
acknowledged.


\end{document}